\documentclass[aps,pra,reprint,nofootinbib,twocolumn,superscriptaddress,showpacs,showkeys,longbibliography]{revtex4-1}

\usepackage{eurosym}
\usepackage{amsmath,amssymb,amstext}
\usepackage{ulem}
\usepackage[usenames,dvipsnames]{color}
\usepackage{graphicx}
\usepackage{braket}
\usepackage{natbib}
\usepackage{comment}
\usepackage{dcolumn}
\usepackage[english]{babel}
\usepackage{wasysym}
\usepackage{bm}
\usepackage[colorlinks,bookmarks=false,citecolor=blue,linkcolor=red,urlcolor=blue]{hyperref}

\usepackage{appendix}
\usepackage{CJK}

\begin{document}
\title{Fast enantioconversion of chiral mixtures based on a four-level double-$\Delta$ model}
\author{Chong Ye}
\affiliation{Beijing Computational Science Research Center, Beijing 100193, China}
\author{Quansheng Zhang}
\affiliation{Beijing Computational Science Research Center, Beijing 100193, China}
\author{Yu-Yuan Chen}
\affiliation{Beijing Computational Science Research Center, Beijing 100193, China}
\author{Yong Li}\email{liyong@csrc.ac.cn}
\affiliation{Beijing Computational Science Research Center, Beijing 100193, China}
\affiliation{Synergetic Innovation Center for Quantum Effects and Applications, Hunan Normal University, Changsha 410081, China}

\begin{abstract}
  Based on the four-level double-$\Delta$ model composed of two degenerated (left- and right-handed) chiral ground states and two achiral excited states, we propose a purely coherent-operation method for enantioconversion of chiral mixtures. By choosing appropriate parameters, the original four-level model will be simplified to two effective two-level sub-systems with each of them involving one chiral ground state.
  Then, with the help of well-designed coherent operations, the initial unwanted and
  wanted chiral ground states are converted, respectively, to the wanted chiral ground state and an auxiliary chiral excited state with the wanted chirality, i.e., achieving the enantioconversion of chiral mixtures. {Comparing with the original works of enantioconversion based on the four-level double-$\Delta$ model with the requirement of the time-consuming relaxation step and repeated operations, our method can be three orders of magnitude faster} since we use only purely coherent operations. Thus, our method offers a promising candidate for fast enantioconversion of chiral molecules with short enantiomeric lifetime or in the experimental conditions where the operation time is limited.

\end{abstract}
\date{\today}
\maketitle
\section{Introduction}
Molecular chirality is extremely important since the vast majority of chemical~\cite{A1}, biological~\cite{A2,A3,A4}, and pharmaceutical~\cite{A5,A7,A8,A6} processes are chirality-dependent.
Enantiodiscrimination, enantioseparation, {enantio-specific state-transfer}, and enantioconversion of chiral mixtures are intensively studied not only in physical chemistry~\cite{PC0,PC1,PC2,PC3,PC4,PC5,Book1}  but also in atomic, molecular, and optical physics~\cite{PRL.87.183002,PRL.99.130403,PRA.77.015403,
JCP.132.194315,JPB.43.185402,JCP.137.044313,
PRA.97.033403,PRA.98.063401,PRL.121.253201,
PRL.120.083204,PRA.99.012513,PRL.122.013204,
PRL.122.024301,PRL.122.173202,PRL.122.223201,
PRX.9.031004,JCP.151.014302,PRA.100.033411,
PRA.100.043403,PRL.121.173002,Nature.497.475,PCCP.16.11114,ACI,JCP.142.214201,JPCL.6.196,JPCL.7.341,
PRL.118.123002,Angew.Chem.56.12512,CA1,CA2,CA3,PRA.84.053849,PRL.118.193401,PRA.93.032508,
PRA.100.043413,PRL.110.213004,PRL.111.023008,PRL.123.243202,PRL.117.033001,PRL.90.033001,PRL.84.1669,
PRA.65.015401,JCP.115.5349,JPB.37.2811}.
Well-established optical methods~\cite{CA1,CA2,CA3}, such as circular dichroism, vibrational circular dichroism, and Raman optical activity, were used for determination of enantiomeric excess in chiral mixtures (enantiodiscrimination). Based on the three-level $\Delta$-type (also called cyclic three-level) models, theoretical methods for enantiodiscrimination~\cite{PRA.100.033411,PRA.84.053849}, {enantio-specific state-transfer}~\cite{JPB.43.185402,PRL.87.183002,PRA.77.015403,
PRL.122.173202,PRA.100.043403,PRA.100.043413}, and enantioseparation~\cite{PRL.99.130403,JCP.132.194315} were proposed. Very recently, breakthrough experiments devoted to realizing enantiodiscrimination~\cite{Nature.497.475,PRL.111.023008,PCCP.16.11114,ACI,JCP.142.214201,JPCL.6.196,JPCL.7.341} and {enantio-specific state-transfer}~\cite{PRL.118.123002,Angew.Chem.56.12512} have been achieved based on the three-level $\Delta$-type models.

Over years, laser-assisted enantioconversion~\cite{PRL.90.033001,PRL.84.1669,PRA.65.015401,JCP.115.5349,JPB.37.2811} has also been investigated theoretically. It is devoted to converting chiral mixtures to enantiopure samples with the help of coherent operations. The unwanted enantiomer, which may be inefficient~\cite{A6} or even cause serious side effects~\cite{A7,A8} in pharmacology, is converted to the wanted enantiomer. Such a promising feature makes
the laser-assisted enantioconversion to be a more ambitious issue related to chiral molecules.
Previously, the four-level double-$\Delta$ model of chiral molecules~\cite{PRL.84.1669,PRA.65.015401,JCP.115.5349,JPB.37.2811} has been introduced to this issue.
Two achiral excited states and each of two
degenerated chiral ground states are coupled by three electromagnetic fields in a form of $\Delta$-type transitions.
The sign of the product of coupling strengths in each $\Delta$-type transition
changes with the chirality of the corresponding chiral ground state.
This fact can evoke a chirality-dependent excitation~\cite{PRA.65.015401}, i.e.,
the molecules initially in the two degenerated chiral ground states are excited differently. {Moreover, the four-level double-$\Delta$ model offers the possibility of converting one chiral ground state to another, which is prohibited in the three-level
$\Delta$-type models with two chiral ground states evolving in separated Hilbert spaces~\cite{PRL.87.183002}.}

{Based on these features of four-level double-$\Delta$ model, the laser-distillation method~\cite{PRL.84.1669,PRA.65.015401,JCP.115.5349,JPB.37.2811} was theoretically proposed
for the laser-assisted enantioconversion by repeating a pair of chirality-dependent excitation and chirality-independent relaxation. It takes about a millisecond for the two achiral excited states relaxing back to the chiral ground states~\cite{JPB.37.2811}. The requirement of repeated operations will make the laser-distillation method~\cite{PRL.84.1669,PRA.65.015401,JCP.115.5349,JPB.37.2811} even more time-consuming.
As a result, the laser-distillation method~\cite{PRL.84.1669,PRA.65.015401,JCP.115.5349,JPB.37.2811} is
unavailable for chiral molecules with short enantiomeric lifetime due to the tunneling between two degenerated chiral states~\cite{quack2003,quack2008} or in the experimental conditions with limited operation time~\cite{MP.110.1757}.}


In this paper, we theoretically propose a novel method using three purely coherent operations to
circumvent this problem. Since we do not need the time-consuming relaxation step, our purely coherent-operation method can be much faster than the laser-distillation method~\cite{PRL.84.1669,PRA.65.015401,JCP.115.5349,JPB.37.2811}.
In our method, by well designing the three electromagnetic fields, the original four-level double-$\Delta$ model is simplified to two effective two-level sub-systems with each of them including one chiral ground state and one of two mutual orthogonal superposition states of the two achiral excited states. The dynamics of the two chiral ground states are subjected to the same effective coupling strengths but different effective detunings in their own two-dimensional subspaces. These properties play center roles in our purely coherent-operation method for enantioconversion.

\section{Simplification of four-level double-$\Delta$ model}\label{SFDM}
The considered four-level double-$\Delta$ model of chiral molecules~\cite{PRL.84.1669,PRA.65.015401,JCP.115.5349,JPB.37.2811} is shown Fig.~\ref{Fig1}(a). The energies of the four states are $\hbar\omega_{A}>\hbar\omega_{S}>\hbar\omega_{R}=\hbar\omega_{L}=0$.  Here, we have assumed the two chiral ground states are degenerate ($\omega_{R}=\omega_{L}$) by neglecting parity violating energy differences due to the fundamental weak force.
{All the four states are chosen as {ro-vibrational states corresponding to the ground electronic state}. With this, disruptive competing processes in the original model~\cite{PRL.84.1669,PRA.65.015401,JCP.115.5349,JPB.37.2811} due to the requirement {of the excited electronic state}, such as dissociation and internal conversion~\cite{PRL.84.1669}, can be avoided.}

{For a chiral molecule, the energy potential surface of the ground electronic state for the vibrational degrees of freedom corresponding to its chirality is of double-well type.
The vibrational sub-levels of the two degenerated chiral states $|L\rangle$ and $|R\rangle$ are located in two bottoms of the double-well energy potential surface, respectively. The vibrational sub-levels of the
pair of achiral states (symmetric one $|S\rangle$ and asymmetric one $|A\rangle$) with a distinguishable vibrational energy difference (e.g. $>2\pi\times10$\,GHz~\cite{JCP.119.5105}) are near or beyond
the barrier of the double-well energy potential surface.}

Three electromagnetic fields with frequencies $\omega_2$, $\omega_0$, and $\omega_1$ are applied to couple the four states in the form of double $\Delta$-type electric-dipole transitions $|Q\rangle\leftrightarrow|A\rangle\leftrightarrow|S\rangle\leftrightarrow|Q\rangle$ with $Q=L,R$.
{The rotational sub-levels of the four states as well as the polarizations, intensities, and frequencies of the three electromagnetic fields are well chosen so that
{the effect of the (electric-dipole) selection-rule allowed transitions out of the four-level double-$\Delta$
model is negligible following Refs.~\cite{JCP.151.014302,PRA.98.063401}. {Specifically, we can choose the rotational sub-levels of $|L\rangle$, $|R\rangle$, $|A\rangle$, and $|S\rangle$ to be
$|J_{\mathrm{k_a k_c M}}=0_{000}\rangle$,
$|0_{000}\rangle$,
$|1_{010}\rangle$, and
$(|1_{101}\rangle+|1_{10-1}\rangle)/\sqrt{2}$ with the $|J_{\mathrm{k_a k_c M}}\rangle$ being the eigenstates of asymmetry top~\cite{PRA.98.063401,JCP.151.014302}. In Fig.~\ref{Fig1}\,(a), the polarizations of the electromagnetic fields are denoted by the colors of the arrowed lines (red, Z-polarized), (blue, Y-polarized), and (black, X-polarized).} In our treatment, for simplicity, we have assumed that~\cite{PRL.84.1669,PRA.65.015401,JCP.115.5349,JPB.37.2811} the coupling {among vibrational, rotational, and electronic states under field-free conditions and tunneling between enantiomers can be neglected. We also have assumed that~\cite{PRL.84.1669,PRA.65.015401,JCP.115.5349,JPB.37.2811} the molecules are in the adiabatic electronic ground state throughout.}

\begin{figure}[h]
  \centering
  \includegraphics[width=0.9\columnwidth]{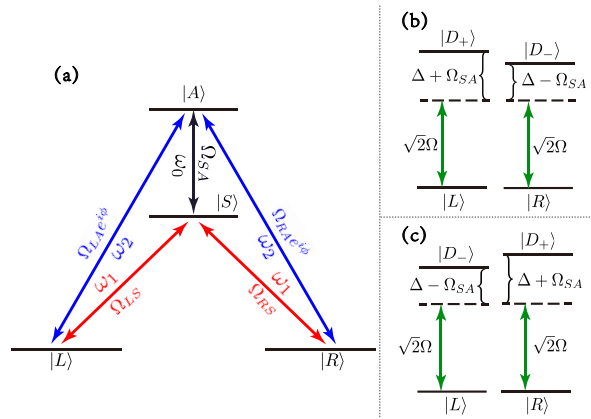}\\
  \caption{(a) The schematic of four-level double-$\Delta$ model of chiral molecules. The chiral ground state $|Q\rangle$ ($Q=L,R$), the asymmetric achiral excited state $|A\rangle$, and the symmetric achiral excited state $|S\rangle$ are coupled in $\Delta$-type transitions $|Q\rangle\leftrightarrow|A\rangle\leftrightarrow|S\rangle\leftrightarrow|Q\rangle$ by three electromagnetic fields with frequencies $\omega_{2}$, $\omega_{0}$, and $\omega_{1}$, respectively. Here $\Omega_{jk}$ ($j,k=L,R,A,S$) are the corresponding (real) coupling strengths and satisfy $\Omega_{LS}=\Omega_{RS}>0$, $\Omega_{LA}=-\Omega_{RA}>0$ and $\Omega_{SA}>0$; $\phi$ is the overall phase of the coupling strengths in the $\Delta$-type transitions related to the left-handed chiral ground state $|L\rangle$. By assuming the three-photon condition of the $\Delta$-type transitions, the one-photon resonance condition of the transition $|A\rangle\leftrightarrow|S\rangle$ (that means $\Delta_{A}=\Delta_{S}\equiv\Delta$ with $\Delta_{A}$ and $\Delta_{S}$ being the detunings corresponding to the transitions $|Q\rangle\leftrightarrow|A\rangle$ and $|Q\rangle\leftrightarrow|S\rangle$), and $\Omega_{LS}=\Omega_{LA}\equiv\Omega$, the original model in (a) can be simplified to two effective two-level sub-systems when the overall phase of the coupling strengths in the $\Delta$-type transitions is well adjusted as (b) $\phi=0$ or (c) $\phi=\pi$.
  }\label{Fig1}
\end{figure}

We are interested in {reducing the original four-level double-$\Delta$ model to the simplified four-level model composed of two separated two-level sub-systems} as shown in Figs.~\ref{Fig1}(b) or \ref{Fig1}(c) by selecting appropriate parameters of the electromagnetic fields.
The frequencies of the three electromagnetic fields should
be chosen following the three-photon resonance condition and the one-photon resonance condition
of the transition $|S\rangle\leftrightarrow|A\rangle$ as
\begin{align}\label{CDT1}
\omega_{2}=\omega_{1}+\omega_{0},~~\omega_{0}=\omega_{A}-\omega_{S}.
\end{align}
Under condition~(\ref{CDT1}), the detunings $\Delta_S\equiv \omega_S-\omega_1$ and
$\Delta_A\equiv \omega_A-\omega_2$ correspond to the transitions $|Q\rangle\leftrightarrow|S\rangle$ and $|Q\rangle\leftrightarrow|A\rangle$ are the same ($\Delta_{A}=\Delta_{S}\equiv\Delta$).
Under the rotating-wave approximation, the system can be
described by the following Hamiltonian in the interaction picture with respect to $\hat{H}_{0}= \omega_{2}|A\rangle\langle A|+\omega_1|S\rangle\langle S|$ as~($\hbar=1$)
\begin{align}\label{HT}
\hat{H}=&\Delta(|A\rangle\langle A|+|S\rangle\langle S|)+(\Omega_{SA}|S\rangle\langle A|+\mathrm{H.c.})\nonumber\\
+&\sum_{Q=L,R}(\Omega_{QA}e^{i\phi}|Q\rangle\langle A|+\Omega_{QS}|Q\rangle\langle S|+\mathrm{H.c.}).
\end{align}
The symmetry breaking related to molecular chirality is reflected in the coupling strengths~\cite{PRL.84.1669,PRA.65.015401,JCP.115.5349,JPB.37.2811} with $\Omega_{LS}=\Omega_{RS}$ and $\Omega_{LA}=-\Omega_{RA}$.
Without loss of generality, we assume $\Omega_{LA}$, $\Omega_{LS}$, and $\Omega_{SA}$ are positive.
Here $\phi$ is the overall phase of the $\Delta$-type transitions $|L\rangle\leftrightarrow|A\rangle\leftrightarrow|S\rangle\leftrightarrow|L\rangle$.
It can be determined by modifying the phases of the three electromagnetic fields.

Moreover, the intensities of the three electromagnetic fields should be adjusted so that
the coupling strengths of the transitions satisfy
\begin{align}
\Omega_{LS}=\Omega_{LA}\equiv\Omega>0.
\end{align}
When the overall phase is tuned to be
$\phi=0$, the dynamics of the two chiral ground states are described by the two-level sub-systems
with Hamiltonian
\begin{subequations}
\begin{align}
\hat{H}_{L}=&(\sqrt{2}\Omega|L\rangle\langle D_{+}|+\mathrm{H.c.})+\Delta_{+}|D_{+}\rangle\langle D_{+}|,\\
\hat{H}_{R}=&(\sqrt{2}\Omega|R\rangle\langle D_{-}|+\mathrm{H.c.})+\Delta_{-}|D_{-}\rangle\langle D_{-}|.
\end{align}
\end{subequations}
The dressed states and the corresponding detunings are
$|D_{\pm}\rangle=(|S\rangle\pm|A\rangle)/{\sqrt{2}}$ and $\Delta_{\pm}=\Delta\pm\Omega_{SA}$.
The original four-level double-$\Delta$ model is simplified to two separated effective two-level sub-systems in the basis $\{|L\rangle,|D_{+}\rangle\}$ and $\{|R\rangle,|D_{-}\rangle\}$ as shown in Fig.~\ref{Fig1}(b).

Similarly, when $\phi=\pi$, the original four-level double-$\Delta$ model can also be simplified to the two separated effective two-level sub-systems with the basis $\{|L\rangle,|D_{-}\rangle\}$ and  $\{|R\rangle,|D_{+}\rangle\}$, respectively, as shown in Fig.~\ref{Fig1}(c). The two-level sub-systems for the two chiral ground states are, respectively, described by $\hat{H}^{\prime}_{L}=(\sqrt{2}\Omega|L\rangle\langle D_{-}|+\mathrm{H.c.})+\Delta_{-}|D_{-}\rangle\langle D_{-}|$ and
$\hat{H}^{\prime}_{R}=(\sqrt{2}\Omega|R\rangle\langle D_{+}|+\mathrm{H.c.})+\Delta_{+}|D_{+}\rangle\langle D_{+}|$.

{With our simplified models of two-level sub-systems as shown in Figs.~\ref{Fig1}(b) and \ref{Fig1}(c), the dynamics of the two chiral ground states are constrained in their own subspaces and are subject to the same coupling strengths but different detunings. These features clearly demonstrate the left-right symmetry breaking in the dynamics of chiral molecules.} Moveover, by changing the overall phase $\phi$ from
$0$ to $\pi$, the subspaces for the two sub-systems change from $\{|L\rangle,|D_{+}\rangle\}$ and $\{|R\rangle,|D_{-}\rangle\}$ to $\{|L\rangle,|D_{-}\rangle\}$ and  $\{|R\rangle,|D_{+}\rangle\}$, which offers the possibility of converting one chiral ground state to another. {We note that such a simplification of the original four-level double-$\Delta$ model can also be used to improve the laser-distillation method for complete enantioconversion~\cite{arXiv.xxx}. In Ref.~\cite{arXiv.xxx}, specific issues of laser-distillation method, which are important but not involved in the original works of laser-distillation method~\cite{PRL.84.1669,PRA.65.015401,JCP.115.5349,JPB.37.2811}, have been discussed. In current work, we will use such a simplification to propose a very different method from the laser-distillation method~\cite{PRL.84.1669,PRA.65.015401,JCP.115.5349,JPB.37.2811,arXiv.xxx}. }

\section{Coherent-operation method}
{Following these properties}, we now present our coherent-operation method for enantioconversion of a racemic mixture with each molecule initially depicted by the density matrix
$\hat{\rho}_{0}=(|L\rangle\langle L|+|R\rangle\langle R|)/2$.
Our method includes three purely coherent operations. For simplicity, we assume
that the coupling strengths and the detunings are time-independent in each operation.
{For the cases of time-dependent coupling strengths, our method can also
be used for enantioconversion by choosing the appropriate parameters (more details see Appendix~\ref{SS2}). }

\begin{figure}[h]
  \centering
  \includegraphics[width=0.9\columnwidth]{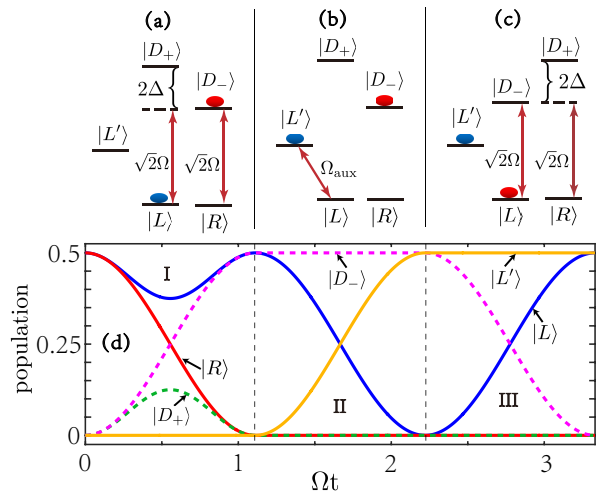}\\
  \caption{coherent-operation method of enantioconversion by simplifying the original four-level double-$\Delta$ model to two separated two-level sub-systems and introducing an auxiliary state $|L^{\prime}\rangle$. (a,b,c) denote, respectively, the first, second, and third coherent operations, where the ellipses stand for the occupied states at the end of each operation in the initial racemic sample with each molecule described by $\hat{\rho}_{0}=(|L\rangle\langle L|+|R\rangle\langle R|)/2$. The corresponding evolution of the population in each state is plotted in panel (d). The dashed lines divide the evolution into three regions labeled with I, II, and III, which correspond to the operations in panels (a), (b), and (c), respectively. {Here the electromagnetic fields applied in panels (a) and (c) are respectively similar to those in Fig.~\ref{Fig1}(b) [under the condition (9)] and Fig.~\ref{Fig1}(c) with $\Delta=\Omega_{SA}$ [under the condition (12)]}. The other parameters are chosen to be $\Omega_{\mathrm{aux}}=\sqrt{2}\Omega$ and $\Delta=\sqrt{6}\Omega$. }\label{Fig4}
\end{figure}

In the first coherent operation, we apply three electromagnetic fields to realize the {simplified four-level model} as shown in Fig.~\ref{Fig4}(a) with the overall phase $\phi=0$ [similar to Fig.~\ref{Fig1}(b)]. We aim to achieve that the wanted chiral ground state $|L\rangle$ evolves back to the state $|L\rangle$ and in the meanwhile the unwanted chiral ground state $|R\rangle$ is transferred to the dressed state $|D_{-}\rangle$ during the first operation. That is, the initial density matrix of each molecule, $\hat{\rho}_{0}$, becomes
$\hat{\rho}_{1}=(|L\rangle\langle L|+|D_{-}\rangle\langle D_{-}|)/2$ accordingly.

For this, we assume that the coupling strengths and the detunings satisfy the following conditions~\cite{PRA.100.043403}
\begin{align}\label{CD1}
&\Omega_{SA}=\Delta=\frac{\sqrt{8n_{L}^2-2(2n_{R}+1)^2}}{2n_{R}+1}\Omega
\end{align}
with integers $n_{L}>n_{R}\ge0$. The operation ends when the chiral ground state $|L\rangle$ evolves
back to itself by experiencing integer $n_{L}$ periods of its corresponding Rabi oscillation.
In the meanwhile, the chiral ground state $|R\rangle$ experiences half-integer $(n_{R}+1/2)$ periods of its corresponding on-resonance Rabi oscillation and evolves to the state $|D_{-}\rangle$. Thus, by applying such an operation to the initial racemic mixture, we can achieve the finial state of $\hat{\rho}_{1}$.

In the second coherent operation, we turn off the two electromagnetic fields coupling with the transitions $|Q\rangle\leftrightarrow|S\rangle$ and $|Q\rangle\leftrightarrow|A\rangle$, keep the electromagnetic field
coupling with transition $|A\rangle\leftrightarrow|S\rangle$, and apply an auxiliary electromagnetic field to resonantly couple with the transition between $|L\rangle$ and a corresponding auxiliary chiral excited state with the same chirality, $|L^{\prime}\rangle$, as shown in Fig.~\ref{Fig4}(b). The working Hamiltonian in the interaction picture under the rotating-wave approximation
\begin{align}\label{Haux}
\hat{H}_{\rm{aux}}&=(\Omega_{\mathrm{aux}}|L^{\prime}\rangle\langle L|+\mathrm{H.c.})+2{\Omega_{SA}}|D_{+}\rangle\langle D_{+}|.
\end{align}
In this case, the wanted chiral ground state $|L\rangle$ can evolve to the axillary chiral excited state of the wanted chirality $|L^{\prime}\rangle$, e.g. experiencing half-integer periods of the Rabi oscillation governed by the following effective on-resonance two-level Hamiltonian $(\Omega_{\mathrm{aux}}|L^{\prime}\rangle\langle L|+\mathrm{H.c.})$. Meanwhile, the dressed state $|D_{-}\rangle$ remains unchanged as indicated by Eq.~(\ref{Haux}).
Thus, at the end of the second operation, the density matrix of each molecule becomes
$\hat{\rho}_{2}=(|L^{\prime}\rangle\langle L^{\prime}|+|D_{-}\rangle\langle D_{-}|)/2$.

In the third coherent operation, we turn off the auxiliary electromagnetic field and construct the {simplified four-level model composed of two separated two-level sub-systems} as shown in Fig.~\ref{Fig4}(c) with the overall phase  $\phi=\pi$ [similar to Fig.~\ref{Fig1}(c)] by applying three electromagnetic fields. By taking the resonant condition in the two-level sub-system including the states $|L\rangle$ and $|D_{-}\rangle$
\begin{align}
\Omega_{SA}=\Delta,
\end{align}
the occupied dressed state $|D_{-}\rangle$ can evolve to the wanted chiral ground state $|L\rangle$ by experiencing half-integer periods of its corresponding on-resonance Rabi oscillation. Thus, each molecule in the initial chiral mixture is converted to the one with the same chirality through
\begin{align}
\hat{\rho}_{0}=\frac{1}{2}(|L\rangle\langle L|+|R\rangle\langle R|)\rightarrow\hat{\rho}_{3}=\frac{1}{2}(|L^{\prime}\rangle\langle L^{\prime}|+|L\rangle\langle L|).
\end{align}

In Fig.~\ref{Fig4}(d), we give an example of our coherent-operation method for converting a racemic sample to an enantiopure sample of the wanted left-handed chirality by choosing $\Omega_{\mathrm{aux}}=\sqrt{2}\Omega$ and $\Delta=\sqrt{6}\Omega$. 
We note that in each operation the (effective) coupling strengths can be chosen independently. Alternatively, when the right-handed chirality is wanted, we can realize the
enantioconversion with the similar process as shown in Fig.~\ref{Fig4} by replacing the overall phases $\phi$ in the first and third coherent operations with $\pi$ and $0$, respectively, and replacing the auxiliary state $|L^{\prime}\rangle$ by an auxiliary one $|R^{\prime}\rangle$. Then, each molecule will be finally depicted with $\hat\rho^{\prime}_{3}=(|R^{\prime}\rangle\langle R^{\prime}|+|R\rangle\langle R|)/2$.

\section{Feasibility and advantage of our method}
Now, we have obtained the enantioconversion {with our coherent-operation method by well designing the electromagnetic fields at each of the three operations based on the four-level double-$\Delta$ model of chiral molecules. In our four-level double-$\Delta$ model and previous ones of others~\cite{PRL.84.1669,PRA.65.015401,JCP.115.5349,JPB.37.2811}, the couplings among vibrational, rotational, and electronic degrees of freedom under field-free conditions and the tunneling between the two enantiomers are assumed negligible. There are evidences~\cite{JCP.129.154314,CJP,JMS} that such an assumption is appropriate for some chiral molecules and inappropriate for others. For the latter chiral molecules where the assumption is not appropriate, the present four-level double-$\Delta$ model will be inadequate to describe their dynamics.}

{For the chiral molecules where the assumption is appropriate,} our method works in the region where the typical {rotational} energy spacing (about $2\pi\times1$\,GHz) of chiral molecules is much larger than the decoherence rate (about $2\pi\times0.1$\,MHz) in current gas-phase experimental conditions~\cite{Nature.497.475,PRL.111.023008,PCCP.16.11114,ACI,JCP.142.214201,JPCL.6.196,JPCL.7.341,
PRL.118.123002,Angew.Chem.56.12512}. In such a region, the intensities of the applied electromagnetic fields can be well chosen so that the decoherence is negligible {compared} with the typical coupling strengths (about $2\pi\times10$\,MHz~\cite{JCP.151.014302,PRL.122.173202}) and the effect of the (electric-dipole) selection-rule allowed transitions out of the four-level double-$\Delta$ model {on the dynamics of molecules initially prepared in the four-state Hilbert space} is negligible. {In addition, we note that the strengths of chemical bonds also set up limitations for intensities of the electromagnetic fields and thus hinder the applications of our method for some chiral molecules.}


{When the coupling strengths are set as about $2\pi\times10$\,MHz, the time cost of our method will be about $0.1$\,$\mathrm{\mu s}$.
That means our method will be much faster than the laser-distillation method~\cite{PRL.84.1669,PRA.65.015401,JCP.115.5349,JPB.37.2811}, where the time cost
is mainly determined by the time-consuming relaxation from the excited achiral states to
the chiral ground states. In current gas-phase experimental conditions~\cite{Nature.497.475,PRL.111.023008,PCCP.16.11114,ACI,JCP.142.214201,JPCL.6.196,JPCL.7.341,
PRL.118.123002,Angew.Chem.56.12512}, the typical relaxation will take more than $10$\,$\mathrm{\mu s}$. Considering
the requirement of repeated operations, the time cost of the laser-distillation method~\cite{PRL.84.1669,PRA.65.015401,JCP.115.5349,JPB.37.2811} will be more than $100$\,$\mathrm{\mu s}$
in the gas-phase experimental conditions~\cite{Nature.497.475,PRL.111.023008,PCCP.16.11114,ACI,JCP.142.214201,JPCL.6.196,JPCL.7.341,
PRL.118.123002,Angew.Chem.56.12512}. Accordingly, our coherent-operation method can be three order faster
than the laser-distillation method~\cite{PRL.84.1669,PRA.65.015401,JCP.115.5349,JPB.37.2811} in the gas-phase experimental conditions~\cite{Nature.497.475,PRL.111.023008,PCCP.16.11114,ACI,JCP.142.214201,JPCL.6.196,JPCL.7.341,
PRL.118.123002,Angew.Chem.56.12512}.}

{The fast feature of our coherent-operation method can be essential for enantioconversion
of chiral mixtures. {For the chiral molecules with short enantiomeric lifetime~\cite{quack2003,quack2008}, our coherent-operation method with a time cost about $0.1$\,$\mathrm{\mu s}$ offers a possible way for enantioconversion, while the laser-distillation method~\cite{PRL.84.1669,PRA.65.015401,JCP.115.5349,JPB.37.2811} may be
unavailable due to the long time cost more than $100$\,$\mathrm{\mu s}$.} For the chiral molecules with long enantiomeric lifetime, our method offers an available candidate
when the operation time is limited by the experimental conditions. In current gas-phase experimental conditions~\cite{Nature.497.475,PRL.111.023008,PCCP.16.11114,ACI,JCP.142.214201,JPCL.6.196,JPCL.7.341,
PRL.118.123002,Angew.Chem.56.12512}, the diffusion of chiral molecules in the apparatus {gives rise to} a limitation for the operation time, out of which the chiral molecules will stick to the apparatus walls and lost~\cite{MP.110.1757}. {We would like to note that our method is not available for chiral molecules with ultrashort enantiomeric lifetime and/or small rotational energy spacings. For these chiral molecules, further methods for enantiodiscrimination, enantioseparation, and enantioconversion are highly desired~\cite{PRL.117.033001}.}  }

\section{Conclusions}
In conclusion, we have proposed theoretically the coherent-operation method for enantioconversion by simplifying the original four-level double-$\Delta$ model to
two separated effective two-level sub-systems. With
three well-designed purely coherent operations, the initial chiral mixture can be converted to an enantiopure sample of the wanted chirality. Specifically, the initial unwanted chiral ground state is converted to the wanted chiral ground state and in the meanwhile the initial wanted chiral ground state is transferred to the auxiliary chiral excited state of the wanted chirality. Our purely coherent-operation method can work much faster than the laser-distillation method~\cite{PRL.84.1669,PRA.65.015401,JCP.115.5349,JPB.37.2811}. {This advantage is essential for the experimental realization of enantioconversion when the operation time is limited due to the short enantiomeric lifetime of the chiral molecules and/or the experimental conditions.}

{Besides the limitation in the operation time, there are other limitations (e.g. phase mismatching due to the spatial distribution of chiral molecules, the Doppler broadening due to the movement of chiral molecules, and the ability of good control of the electromagnetic fields) hindering experimental efforts to achieve the enantioconversion as well as enantiodiscrimination and {enantio-specific state-transfer} based on few-level models.
With the development of experimental technologies, such as further cooling the chiral molecules and using more advanced IR or microwave techniques, the effects of these limitations can be overcome or suppressed. }

\section*{Acknowledgement}
This work was supported by the National Key R\&D Program of China grant (2016YFA0301200), the Natural Science Foundation of China (under Grants No.~11774024, No.~11534002, No.~U1930402 and No. 11947206), and the Science Challenge Project (under Grant No.~TZ2018003).

\appendix
\section{Rotational sub-levels of the working states}\label{SS1}
{For the chiral molecules where the couplings among vibrational, rotational, and electronic degrees of freedom under field-free conditions and the tunneling between the two enantiomers
are negligible, the rotations of molecules can be introduced by considering the rotational sub-levels of
the working states as in Sec.~\ref{SFDM}. Usually, including the rotational sub-levels will
challenge the few-level model due to the magnetic degeneracy and the small rotational energy spacings~\cite{JCP.137.044313}. Specifically, other ro-vibrational transitions out of the few-level model
are usually connected to the working states in the few-level model. For three-level cyclic models, researchers have pointed
out the effects of these transitions are negligible by well designing the electromagnetic fields due to the
selection rules of electric-dipole transition and/or large-detuning approximation~\cite{JCP.151.014302,PRA.98.063401}. In the following, we will focus on the similar
issue in our four-level double-$\Delta$ models to show the transitions out of our model are negligible
by well designing the electromagnetic fields. This will eventually protect the validity of the four-level double-$\Delta$ models of chiral molecules where the couplings among vibrational, rotational, and electronic degrees of freedom
under field-free conditions and the tunneling between the
two enantiomers are negligible.
For this purpose, we would like to use HSOH molecule, whose ro-vibrational spectrum and transition
electric dipoles are accessible~\cite{JMS.257.57}, as a toy model without considering the couplings among vibrational, rotational, and electronic degrees of freedom under field-free conditions and the tunneling between the two enantiomers~\cite{PRL.117.033001,JCP.151.014302,PRL.120.083204}.}

\begin{figure}[h]
  \centering
  \includegraphics[width=0.9\columnwidth]{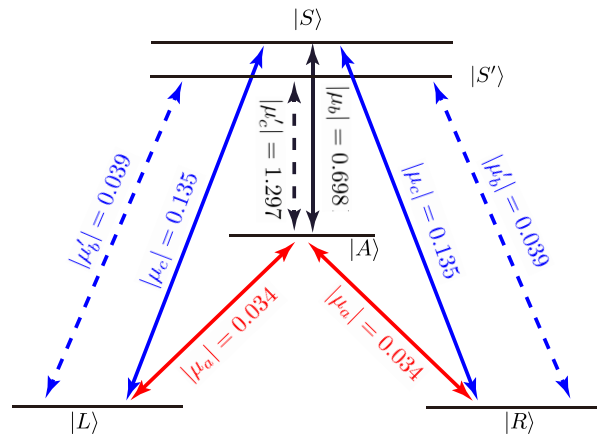}\\
  \caption{Four-level double-$\Delta$ model of HSOH molecule (related
  transitions denoted by solid arrowed lines. The ro-vibrational states denoted by solid line are $|L\rangle=|0^{L}\rangle\otimes|0_{000}\rangle$,
  $|R\rangle=|0^{R}\rangle\otimes|0_{000}\rangle$,
  $|A\rangle=|2v^{-}_{\mathrm{HSOH}}\rangle\otimes|1_{010}\rangle$, and
  $|S\rangle=|2v^{+}_{\mathrm{HSOH}}\rangle\otimes(|1_{101}\rangle+|1_{10-1}\rangle)/\sqrt{2}$. The energies of $|Q\rangle$, $|S\rangle$, and $|A\rangle$ are
0\,GHz, $2\pi\times25493.7135$\,GHz, and $2\pi\times25400.5897$\,GHz.
  The transitions denoted by arrowed lines of the same color are coupled by the same electromagnetic field.
  The polarizations of the electromagnetic fields are denoted by the colors of the arrowed lines (red, Z-polarized), (blue, Y-polarized), and (black, X-polarized). We also give the transition dipole in the molecule-frame for each transition. The transitions {with the most possibility} to challenge the four-level double-$\Delta$ model denoted by dashed arrowed lines ($|S^{\prime}\rangle=|2v^{+}_{\mathrm{HSOH}}\rangle\otimes(|1_{111}\rangle+|1_{11-1}\rangle)/\sqrt{2}$).
  The energy of $|S^{\prime}\rangle$ is $2\pi\times25493.2718$\,GHz. They are negligible under the large-detuning condition.
 }\label{FigS1}
\end{figure}

In Fig.~\ref{FigS1}, we show the four-level double-$\Delta$ model of HSOH molecule, where the related
transitions are denoted by solid arrowed lines.
The vibrational sub-levels for the two achiral
states $|S\rangle$ and $|A\rangle$ are chosen
as the second pair of excited states $|2v^{+}_{\mathrm{HSOH}}\rangle$ and $|2v^{-}_{\mathrm{HSOH}}\rangle$ (``$\pm$'' denote the parity under the inversion operation) in the vibrational degrees of freedom $\tau_{\mathrm{HSOH}}$ corresponding to the chirality of HSOH molecule~\cite{JMS.257.57}. The vibrational energies for
the symmetric and asymmetric states are $2\pi\times25276.3629$\,GHz and $2\pi\times25370.4676$\,GHz~\cite{JMS.257.57}, respectively.
The vibrational energy difference between them is $2\pi\times94.1047$\,GHz. The vibrational sub-levels of
the two chiral ground states are $|0^{L}\rangle$ and $|0^{R}\rangle$ are located at the bottoms of
the double-well type potential energy surface for the vibrational degrees
of freedom $\tau_{\mathrm{HSOH}}$, respectively. When the molecular rotations are considered,
we have $|L\rangle=|0^{L}\rangle\otimes|J_{\mathrm{k_a k_c M}}=0_{000}\rangle$,
$|R\rangle=|0^{R}\rangle\otimes|0_{000}\rangle$,
$|A\rangle=|2v^{-}_{\mathrm{HSOH}}\rangle\otimes|1_{010}\rangle$, and
$|S\rangle=|2v^{+}_{\mathrm{HSOH}}\rangle\otimes(|1_{101}\rangle+|1_{10-1}\rangle)/\sqrt{2}$ with the $|J_{\mathrm{k_a k_c M}}\rangle$ notation~\cite{PRA.98.063401,JCP.151.014302}.
Accordingly, the transitions
$|Q\rangle\leftrightarrow|S\rangle$, $|Q\rangle\leftrightarrow|A\rangle$, and $|A\rangle\leftrightarrow|S\rangle$ are c-type, a-type, and b-type~\cite{PRA.98.063401,JCP.151.014302}
with the transition electric dipoles in the molecule-fixed frame as
$|\mu_{c}|=0.135$\,Debye, $|\mu_{a}|=0.034$\,Debye, and $|\mu_{b}|=0.698$\,Debye~\cite{JMS.257.57}.
The rotational constants are $A=2\pi\times202.0687$\,GHz, $B=2\pi\times15.2819$\,GHz,
and $C=2\pi\times14.8402$\,GHz~\cite{JMS.257.57}. The energies of $|Q\rangle$, $|S\rangle$, and $|A\rangle$ are 0\,GHz, $2\pi\times25493.7135$\,GHz, and $2\pi\times25400.5897$\,GHz. We note that the state
$|S\rangle$ has higher energy than $|A\rangle$ due to the contribution of rotational energies, which is different from the case in our main text where $|S\rangle$ has lower energy than $|A\rangle$. But this difference will not change the results of our enantioconversion.

{Usually, the magnetic degeneracy may challenge the feasibility of the four-level $\Delta$ model. Here, we
avoid this problem by appropriately choosing the polarizations of our three applied electromagnetic
fields according to the selection rules of electric-dipole transition. The polarizations are denoted by the colors of the arrowed lines in Fig.~\ref{FigS1} with red for Z-polarized, blue for Y-polarized, and black for X-polarized.  More details can be found in Refs.~\cite{JCP.151.014302,PRA.98.063401}.}
The three electromagnetic fields are nearly-resonantly or resonantly coupled to the chosen transitions
(denoted by solid arrowed lines) in Fig.~\ref{FigS1}.
In our main text, we require that the coupling strength of each
transition is about $2\pi\times10$\,MHz. Compared with the coupling strength,
the typical decoherence rate (about $2\pi\times0.1$\,MHz) in current gas-phase
experimental conditions~\cite{Nature.497.475,PRL.111.023008,PCCP.16.11114,ACI,JCP.142.214201,JPCL.6.196,JPCL.7.341,
PRL.118.123002,Angew.Chem.56.12512} is negligible.
For the transitions
$|Q\rangle\leftrightarrow|S\rangle$, the required IR field
should have the intensity about $10$\,$\mathrm{W/cm^2}$.
For the transitions $|Q\rangle\leftrightarrow|A\rangle$, the required IR field
should have the intensity about $100$\,$\mathrm{W/cm^2}$.
For the transition $|A\rangle\leftrightarrow|S\rangle$, the required microwave
field should have the intensity about $0.2$\,$\mathrm{W/cm^2}$.
These intensities of IR and microwave fields are experimentally
available~\cite{JCP.151.014302}.

{There are still other (electric-dipole) selection-rule allowed transitions may challenge the four-level double-$\Delta$ model due to the other ro-vibrational states whose energies are close to the chosen four states. These transitions are negligible under the large-detuning condition. In Fig.~\ref{FigS1}, we give the transitions with most possibility to challenge our four-level double-$\Delta $ model (denoted by dashed arrowed lines).}
The microwave field corresponding to the transition $|A\rangle\leftrightarrow|S\rangle$ can couple with the c-type transition $|A\rangle\leftrightarrow|S^{\prime}\rangle$ with $|\mu^{\prime}_{c}|=1.297$\,Debye~\cite{JMS.257.57} and $|S^{\prime}\rangle=|2v^{+}_{\mathrm{HSOH}}\rangle\otimes(|1_{111}\rangle+|1_{11-1}\rangle)/\sqrt{2}$.
The energy of $|S^{\prime}\rangle$ is $2\pi\times25493.2718$\,GHz. The energy difference between $|S\rangle$
and $|S^{\prime}\rangle$ is $2\pi\times441.7$\,MHz. Since we require the transition $|A\rangle\leftrightarrow|S\rangle$
is coupled resonantly, the corresponding detuning of the transition $|A\rangle\leftrightarrow|S^{\prime}\rangle$ is $2\pi\times441.7$\,MHz.
Under the required intensity about $0.2$\,$\mathrm{W/cm^2}$,
the coupling strength of the transition $|A\rangle\leftrightarrow|S^{\prime}\rangle$ is about $2\pi\times50$\,MHz, which is much smaller than the corresponding detuning $2\pi\times441.7$\,MHz.
Thus, the transition $|A\rangle\leftrightarrow|S^{\prime}\rangle$ is negligible.

The IR field corresponding to the transitions $|Q\rangle\leftrightarrow|S\rangle$ can couple with
the b-type transitions $|Q\rangle\leftrightarrow|S^{\prime}\rangle$
with $|\mu^{\prime}_{b}|=0.039$\,Debye~\cite{JMS.257.57}. Under the required intensity about $10$\,$\mathrm{W/cm^2}$, the
coupling strengths of the transitions $|Q\rangle\leftrightarrow|S^{\prime}\rangle$ are about
$2\pi\times1$\,MHz. The corresponding detuning of the transitions $|Q\rangle\leftrightarrow|S^{\prime}\rangle$ are about $2\pi\times441.7$\,MHz, which is much larger than the corresponding coupling strength. Thus, the transitions $|Q\rangle\leftrightarrow|S^{\prime}\rangle$ are negligible.

{We note that the working states should be chosen appropriately. For an example, if we take the $|S^{\prime}\rangle$ as the working states in Fig.~\ref{FigS1} instead of $|S\rangle$ to construct an alternative four-level double-$\Delta$ model for enantioconversion, it is unavailable to ensure the
decoherence negligible and the large-detuning limit, simultaneously. Specifically, when the
transition $|Q\rangle\leftrightarrow|S^{\prime}\rangle$ in the alternative four-level double-$\Delta$ model
is coupled with the driven field at the strength of
$2\pi\times10$\,MHz to make the decoherence negligible, the corresponding coupling strength for the transition $|Q\rangle\leftrightarrow|S\rangle$ out of the alternative four-level double-$\Delta$ model driving by the same field will be of order of $2\pi\times100$\,MHz. The transition $|A\rangle\leftrightarrow|S\rangle$ out of the alternative four-level double-$\Delta$ model with the detuning $2\pi\times441.7$\,MHz can not be neglected, since the large-detuning limit is not satisfied.}

\section{Enantioconversion with time-dependent models}\label{SS2}
In the main text, we assume the time-independent model in each of our coherent operations.
In practice, it takes time for the applied electromagnetic fields to reach their required intensities. That is, the envelope of the laser field is time dependent. In this section, we will show our method can still work in the time-dependent case.

We can first turn on the electromagnetic field coupling the transition
$|A\rangle\leftrightarrow|S\rangle$ and maintain its intensity such
that $\Omega_{SA}=-\Delta>0$ before finishing all of our operations.
In the first operation with time duration from $t_0$ to $t_1$, we apply the other two optical fields corresponding to $\Omega~ (=\Omega_{LS}=\Omega_{LA})$.
The evolution matrices of the two initial chiral ground states are
\begin{align}\label{DFU}
&U_{L}\equiv\mathcal{T}\int^{t_1}_{t_0}e^{-i[\sqrt{2}\Omega(t)\sigma^{L}_{x}+\Delta(I^{L}_{0}+\sigma^{L}_{z})]}dt\nonumber\\
&U_{R}\equiv\mathcal{T}\int^{t_1}_{t_0}e^{-i\sqrt{2}\Omega(t)\sigma^{R}_{x}}dt.
\end{align}
{Here, $\sigma^{Q}_{x,y,z}$ are the $2\times2$ pauli matrices and $I^{Q}_{0}$ is the $2\times2$ unit matrix in the corresponding basis of the two chiral ground state $\{|L\rangle,|D_{+}\rangle\}$ and $\{|R\rangle,|D_{-}\rangle\}$.}

The area of the pulses corresponding to
transitions $|Q\rangle\leftrightarrow|A\rangle$ and $|Q\rangle\leftrightarrow|S\rangle$
should be specified as
\begin{align}\label{BZD}
\int^{t_1}_{t_0}\sqrt{2}\Omega(t)dt={(2l+1)}\frac{\pi}{2}
\end{align}
with integer $l$, so that $U_{R}=\pm\sigma^{R}_{x}$ and the right-handed chiral ground state
is transferred to upper-level ($|D_{-}\rangle$) of its corresponding two-level system. By tuning the (time-independent) parameter $\Delta$, the left-handed chiral ground state can (approximately) evolve back to itself (more information can be found in Ref.~\cite{arXiv.xxx}).

In the second and third step, the working (effective) two-level models are on resonance.
The area of the (effective) coupling strengths should be $(2l+1)\pi/2$, which can also be
obtained when the working models are time-dependent.
{In addition, we note that the envelope of each laser field should change much slowly comparing with its corresponding transition frequency and rotational energy spacings. In this way, each laser field will only couple with its corresponding transition and the rotating-wave approximation is valid.}

{}

\end{document}